\documentclass[12pt]{aa}

\usepackage{txfonts}
\usepackage[caption=false]{subfig}

\bibpunct{(}{)}{;}{a}{}{,} 
\usepackage{graphicx}
\usepackage[utf8]{inputenc}
\usepackage[T1]{fontenc}
\usepackage{dcolumn}
\usepackage{bm}
\usepackage{color}
\usepackage{soul}
\usepackage{url}
\usepackage[normalem]{ulem}
\expandafter\let\csname equation*\endcsname\relax
\expandafter\let\csname endequation*\endcsname\relax
\usepackage{amsmath}

\usepackage{siunitx}
\usepackage{aas_macros}
\usepackage{multirow}
\usepackage{float}

\usepackage[caption=false]{subfig}
\usepackage{tikz} 
\makeatletter
\newcommand*{\centerfloat}{%
  \parindent \z@
  \leftskip \z@ \@plus 1fil \@minus \textwidth
  \rightskip\leftskip
  \parfillskip \z@skip}
\makeatother

\tikzset{%
  base/.style = {rectangle, rounded corners, draw=black,
                  minimum width=0.45\columnwidth, minimum height=3.em,
                  text centered, font=\sffamily},
  inpu/.style = {base, fill=blue!15},
  conv/.style = {base, fill=red!15},
  dens/.style = {base, fill=green!15},
  outp/.style = {base, fill=yellow!15},
}

\bibpunct{(}{)}{;}{a}{}{,}

\begin{document}

\title
{LSTM and CNN application for core-collapse \\ supernova search in gravitational wave real data}
\author{Alberto Iess \inst{\ref{inst1}} \and Elena Cuoco \inst{\ref{inst1} \and \ref{inst2}} \and Filip Morawski \inst{\ref{inst3} \and \ref{inst4}}  \and Constantina Nicolaou \inst{\ref{inst5}} \and Ofer Lahav \inst{\ref{inst5}}}

\institute{Scuola Normale Superiore, Piazza dei Cavalieri 7, I-56126 Pisa, Italy \email{alberto.iess@sns.it}
\label{inst1}
\and European Gravitational Observatory (EGO), I-56021 Cascina, Pisa, Italy \label{inst2}
\and Nicolaus Copernicus Astronomical Center, Polish Academy of Sciences, Bartycka 18, 00-716, Warsaw, Poland \label{inst3}
\and SUPA, School of Physics and Astronomy, University of Glasgow, Glasgow G12 8QQ, UK \label{inst4}
\and Department of Physics \& Astronomy, University College London, Gower Street, London, WC1E 6BT, UK \label{inst5}
}

\vspace{10pt}

\abstract{

\textit{Context.} Core-collapse supernovae (CCSNe) are expected to emit gravitational wave signals that could be detected by current and future generation interferometers within the Milky Way and nearby galaxies. The stochastic nature of the signal arising from CCSNe requires alternative detection methods to matched filtering. 

\textit{Aims.} We aim to show the potential of machine learning (ML) for multi-label classification of different CCSNe simulated signals and noise transients using real data. We compared the performance of 1D and 2D convolutional neural networks (CNNs) on single and multiple detector data. For the first time, we tested multi-label classification also with long short-term memory (LSTM) networks.

\textit{Methods.} We applied a search and classification procedure for CCSNe signals, using an event trigger generator, the Wavelet Detection Filter (WDF), coupled with ML. We  used time series and time-frequency representations of the data as inputs to the ML models. To compute classification accuracies, we simultaneously injected, at detectable distance of 1\,kpc, CCSN waveforms, obtained from recent hydrodynamical simulations of neutrino-driven core-collapse, onto interferometer noise from the O2 LIGO and Virgo science run.

\textit{Results.} We compared the performance of the three models on single detector data. We then merged the output of the models for single detector classification of noise and astrophysical transients, obtaining overall accuracies for LIGO ($\sim99\%$) and ($\sim80\%$) for Virgo. We extended our analysis to the multi-detector case using triggers coincident among the three ITFs and achieved an accuracy of $\sim98\%$.}

\titlerunning{LSTM and CNN for CCSNe search in gravitational wave real data}
\authorrunning{Alberto Iess et al.}
\maketitle

\section{Introduction}
\label{sec:intro}
Since the discovery of binary black hole (BBH) merger GW150914 \citep{PhysRevLett.116.061102} by the LIGO and Virgo Scientific Collaboration, gravitational wave (GW) astronomy has gained momentum in the scientific
community by providing a new observational channel in astrophysics. This led to the first joint electromagnetic (EM) and GW multi-messenger observation of a binary neutron star (BNS) merger, GW170817 \citep{PhysRevLett.119.161101}, during the second science run of the Advanced LIGO (\cite{aLIGO} and \cite{aLIGO1}) and Advanced Virgo \citep{AdVirgo} interferometric detectors, along with a large number of BBH mergers collected in the first GW transient catalog GWTC-1 \citep{catalog}. In 2020, LIGO and Virgo completed the third science run O3; results from the first part of the run, O3a, which led to the discovery of
39 candidate binary merger events in 26 weeks of data are presented in the GWTC-2 catalogue \citep{catalog2}. Currently the LIGO, Virgo interferometers are following a schedule which will implement upgrades to increase the sensitivity for the O4 science run, to begin in 2023. The KAGRA detector \citep{KAGRA}, which started its first observing run towards the end of O3 \citep{Akutsu2020}, will also increase its sensitivity effectively producing a network of four interferometers with increased capabilities for future runs.

Among astrophysical sources of GWs, core-collapse supernovae (CCSNe) are yet to be observed and are an interesting candidate for multi-messenger analysis due to their EM and neutrino emission. However GW signals from CCSNe cannot be exactly modelled due to the stochasticity involved in the collapse dynamics and the dependency on
many parameters such as the progenitor mass, rotational state, and metallicity. Therefore alternative detection methods to the usual matched filtering approach are used which make minimal assumptions on the signal waveform, such as the wavelet-based  coherent WaveBurst pipeline (\cite{Klimenko2004} and \cite{Klimenko2016}). Machine learning (ML) techniques have been used to tackle many aspects of GW astronomy including signal detection for different types of sources (\cite{Baker2015}, \cite{Gabbard2017}, \cite{Kim2020}, and \cite{Morawski2020}), parameter estimation (\cite{Varma2019}, \cite{Haegel2020}, \cite{Green2020}, \cite{Chua2020}, and \cite{Williams2021}), noise classification (\cite{Zevin2017}, \cite{Mukund2017}, and \cite{cuoco-glitch}), and subtraction (\cite{Torres2020} and \cite{Wei2020}), as described in the comprehensive review by \cite{MLreview}. In previous work \citep{Iess2020} we showed the efficacy of 1D and 2D convolutional neural networks (CNNs) in distinguishing between noise transients, known as glitches, and waveforms from CCSNe burst signals produced through 3D hydrodynamical simulations and added to background noise, using whitened time series and spectrograms as input features. The background noise was simulated and gaussian, produced from the theoretical sensitivity curves of Virgo during O3 science run and the future Einstein Telescope (ET) detector (\cite{Punturo2010} and \cite{Hild_2011}). In the same work, we demonstrated the capability of correctly classifying among different CCSNe GW emission models. CNNs have also been used by  \cite{2018PhRvD..98l2002A} to search for g-mode emission in CCSNe signals in multiple GW detectors using a phenomenological model. \cite{Cavaglia2019} used a genetic programming algorithm which takes as input features a subset of computed parameters such as central frequency, bandwith and durations to achieve the same goal on single interferometer data. \cite{PhysRevD.102.043022} trained a 1D CNN model to search for CCSNe using the whitened GW time series from a network of GW detectors. 

In this analysis we extended previous work  on the subject \citep{Iess2020} by using real detector noise from the O2 science run instead of simulated noise. We increased the number of supernova waveforms by adding two higher energy models. Moreover we explored recurrent neural networks (RNNs) and compared their performance to CNNs. RNNs are a type of neural network specialised for processing sequential data due to a feedback loop mechanism that allows the network to retain in memory previous inputs and consider these along with the current input \citep{Husken2003}. RNN models are commonly used in the fields of natural language processing \citep{Nicolaou2019} and for time-series data. In this work we used a particular type of RNN, long short-term memory cells (LSTM) \citep{Hochreiter1997}, which provide a solution to the vanishing gradient problem and have the capacity to handle long-term dependencies making them best suited for analysing gravitational wave data.
We started our analysis by training and testing on a piece of data from a single detector and later extended our approach to a three-interferometer network. The paper is structured in the following way: Section \ref{sec:dataset} and \ref{sec:waveforms} provide insight on the real detector data and a brief overview of the waveforms used in this study. Section \ref{sec:ml} describes each of the ML algorithms involved in the analysis. The results are detailed in Section \ref{sec:res}, followed by  discussions and conclusions in Section \ref{sec:disc}.

\section{Dataset}
\label{sec:dataset}
\begin{figure*}[!t]
\begin{minipage}{.6\textwidth}
\centering
    \hspace*{-1cm}
    \includegraphics[width=0.70 \textwidth]{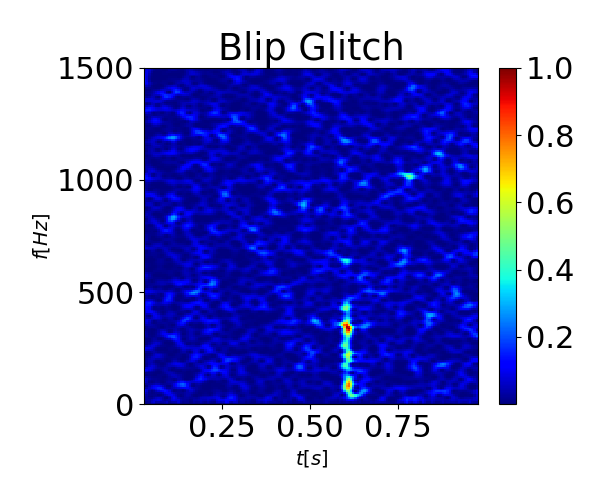}
\end{minipage}
\begin{minipage}{.6\textwidth}
\centering
    \hspace*{-6cm}
    \includegraphics[width=0.70 \textwidth]{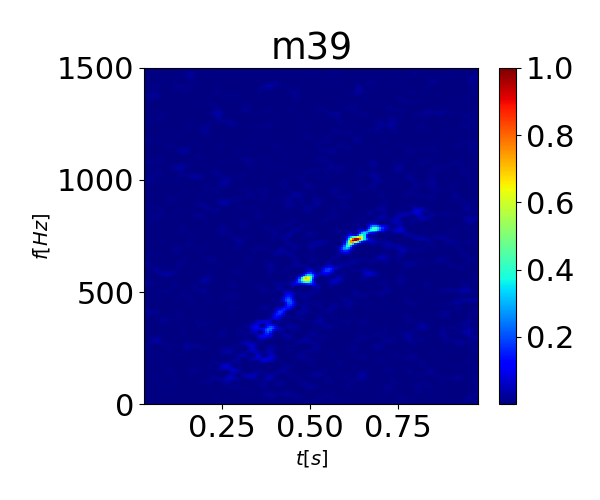}
\end{minipage}
  \caption{Sample whitened spectrograms for the noise and signal classes used in this analysis: a blip glitch (\emph{left}) and m39 (\emph{right}), a rapidly rotating CCSN explosion model. The GW shown for m39 is for emission at the pole. The spectrograms are rescaled to the $[0,1]$ range for a 2D CNN classifier.}
  \label{fig:samples}
\end{figure*}

 The data analysed in this study was generated by adding selected CCSNe waveforms, obtained through 3D numerical simulations, to real interferometer noise. Compared to simulated gaussian noise used in \cite{Iess2020}, \cite{2018PhRvD..98l2002A}, and \cite{PhysRevD.102.043022}, real noise in GW detectors is subject to a certain degree of variability as it depends on environmental conditions which can be monitored but not controlled. When noise artifacts have long timescales they affect the sensitivity of the detector, which in the frequency domain corresponds to a non-stationary noise power spectral density (PSD). Moreover, many classes of short-lived non-astrophysical transients are also present in the data and can hinder GW search pipelines by triggering vetoes or false alarms. Keeping this in mind and in order to compare results, we selected 44 segments from the O2 public science run \citep{theligoscientificcollaboration2019open}, requiring that they pass the data quality flags characterised in \cite{dataquality2016} at least $97\%$ of the time for each detector. Each file contains a $4096$\,s long time series sampled at $4096$\,Hz and is identified by its initial GPS time. The segments span the period from $t_{GPS}=1185669120$ to $t_{GPS}=1186070528$. In this analysis we used five CCSN models previously tested on simulated data (s11, s13, he3.5, s18p, s25). Additionally, we included two higher amplitude models with Wolf-Rayet progenitor stars (m39, y20) and a waveform with the same progenitor mass of s18p, but no initial density perturbations in the convective oxygen shell (s18np) \citep{10.1093/mnras/staa1048}, to increase the complexity of the classification problem. The waveforms were downsampled to $4096$\,Hz, applying a low-pass filter to suppress aliasing. We simulated all signals at fixed distance of $1$\,kpc and uniformly distributed in the sky. To obtain the latter condition, we sampled from a uniform distribution in the right ascension $\alpha$ and the cosine of the declination $\delta$. The polarization angle in the wave frame was fixed at $\psi=0$. In each segment the first 300 seconds do not contain injections and were used to estimate the power spectral density (PSD) \citep{Cuoco2001} and compute the whitening parameters. The signals were then injected every ten seconds taking into account the antenna pattern functions $F_+$, $F_\times$ for the two gravitational wave polarizations $h_+$, $h_\times$ at the different detectors at each epoch, along with the time shift computed for each interferometer with respect to the geocentric frame \citep{schutz2011}. The full dependencies are described in the following formula:
\begin{equation}
    h(t) = F_+(\alpha, \delta, \lambda, \beta, \chi, \eta) \, h_+(t) + F_\times(\alpha, \delta, \lambda, \beta,\chi, \eta) \, h_\times(t), 
\end{equation}
 where $\lambda$, $\beta$ are the longitude and latitude of the detector, $\chi$ defines the orientation of the bisector of the detector arms, and $\eta$ the angle between the two arms. Let $s(t) = n(t) + h(t)$ be the detector time series, with $n(t)$ interferometer noise and $h(t)$ a possible signal. The matched filter signal-to-noise ratio (S/N) time series at time $t_0$ is then computed as in  \cite{Allen2012}:
\begin{equation}
    \left(\frac{S}{N}\right)^2 = 4  \int_0^{f_{max}}{ \frac{\tilde{s}(f) \left[\tilde{h}(f)^*\right]_{t_0 = 0} }{S_n(f)} \, e^{2\pi i f t_0} \, df},
\end{equation}
where $f_{max}$ is the maximum frequency cutoff and ${S_n(f)}$ is the one-sided noise PSD. After injecting the signals into the detector strain data, a whitening step in time domain was applied by means of the Wavelet Detection Filter (WDF) library, documented in \cite{Cuoco2001} and \cite{CuocoRazzano2018}, in order to remove the stationary contribution to the noise PSD. The whitening parameters were computed by fitting the noise PSD using an Auto-Regressive (AR) model. The total number of CCSN injected in each segment is $376$, $47$ for each of the eight models. A subset of the total $16544$ CCSN signals injected in the dataset will trigger the WDF pipeline by means of a fixed threshold set on the squared sum of the wavelet transform coefficients. These will not be the only events detected by WDF, since noise transients provide additional triggers. Whitened spectrograms for instances of these two types of instrumental and astrophysical burst signals are pictured in Figure \ref{fig:samples}.

\section{Waveforms}
\label{sec:waveforms}
\subsection{Model s11}
Model s11 is the $11\,M_{\odot}$ ZAMS model simulated by \cite{2017MNRAS.468.2032A} with the \textsc{PROMETHEUS-VERTEX} code \citep{2002A&A...396..361R}, which employs Newtonian gravity. The simulation ends 0.35\,s after the core bounce time. This model does not explode and has the smallest GW amplitude of all the GW signals considered in this study. It has a lower peak frequency at $\sim600$\,Hz.

\subsection{Model s13}
Model s13 is a $13\,M_{\odot}$ ZAMS model simulated by \cite{2019ApJ...876L...9R} using the Eulerian radiation-hydrodynamics code \textsc{FORNAX} \citep{2019ApJS..241....7S}. This model does not explode, and shows GW emission associated with g-modes. This model ends at 0.78\,s after core bounce. Due to the lack of shock revival, this model has low GW amplitude and peaks at a frequency of $\sim800$\,Hz. This model presents a weak standing accretion shock instability (SASI) component.

\subsection{Model s25}
Model s25 is a $25\,M_{\odot}$ ZAMS model simulated by \cite{2019ApJ...876L...9R} using the Eulerian radiation-hydrodynamics code \textsc{FORNAX}. It explodes at 0.5\,s after core bounce and the simulation ends at 0.62\,s after core bounce time when the GW emission is still high. This model exhibits a clear signature of the SASI at low-frequency as well as high frequency g-modes, with the peak GW emission at $\sim1000$\,Hz.

\subsection{Model s18p}
Model s18p is the $18\,M_{\odot}$ ZAMS progenitor from \cite{2018arXiv181205738P} simulated with the general relativistic neutrino hydrodynamics code \textsc{CoCoNuT-FMT}  \citep{2010ApJS..189..104M}. The simulation end time is 0.9\,s, at which time the GW emission has reached very low amplitudes. This model shows a clear g-mode signal in the spectrogram. This model explodes at $\sim300$\,ms after core bounce. The GW frequency peaks at $\sim850$\,Hz. 

\subsection{Model s18np}
Model s18np from \cite{10.1093/mnras/staa1048} differs from s18 in the fact that the simulation does not include perturbations from the convective oxygen shell. As a result, this model develops strong SASI after collapse.

\subsection{Model he3.5}
Model he3.5 is the $3.5\,M_{\odot}$ ultra-stripped helium star from \cite{2018arXiv181205738P} simulated with the general relativistic neutrino hydrodynamics code \textsc{CoCoNuT-FMT}. An ultra-stripped star is a star in a binary system that has been stripped of it's outer layers due to mass transfer to the binary companion star \citep{2015MNRAS.451.2123T}. The simulation ends at 0.7\,s after core-bounce time, well after the peak GW emission phase. This model shows a clear g-mode in the spectrogram. The amplitude of the GW signal is strongest at $\sim900$\,Hz. This model explodes at $\sim0.4$\,s after core bounce. 

\subsection{Model y20}
Model y20 simulated by \cite{10.1093/mnras/staa1048} with the neutrino hydrodynamics code CoCoNuT-FMT is a non-rotating, 20 solar mass Wolf-Rayet star with solar metallicity. The simulation of this model is stopped at 1.2 s after core bounce. The waveform is characterized by prompt convection after core bounce, below $200$\,Hz. Shock revival occurs $\sim200$\,ms after the bounce and is followed by a strong fundamental mode emission.

\subsection{Model m39}
Model m39 simulated by \cite{10.1093/mnras/staa1048} is a rapidly rotating Wolf-Rayet star with an initial helium star mass of 39 solar masses. The simulation is performed with the neutrino hydrodynamics code CoCoNuT-FMT. A large signal amplitude is produced by the strong neutrino-driven explosion thanks to rapid rotation ($600 \si{\km \s}^{-1}$), without the aid of strong magnetic fields. The simulation ends $0.98$\,s after core bounce and shock revival is achieved $\sim200$\,ms after bounce. The main feature is a a strong fundamental mode emission up to a peak frequency of $\sim800$\,Hz. The equatorial emission from this model has high amplitude over all frequencies associated with core bounce of a rapidly rotating star. However, we only used the polar emission in our analysis, shown in Figure \ref{fig:samples}.

\section{Deep learning algorithms}
\label{sec:ml}
For this analysis we used different types of ML algorithms for multi-label classification. Two types of inputs are provided: the whitened time series for the 1D CNN and the LSTM network, the whitened time-frequency image for the 2D CNN.  

\subsection{CNN 1D}
The 1D CNN takes the whitened timeseries as inputs. It is composed of 4 convolutional layers with kernels of size 3. The number of convolutional filters are (120,80,80,40). Max pooling (2,2) and spatial dropout on $40\%$ of nodes is applied after each convolutional layer. The output from the CNN architecture is fed into two Fully Connected (FC) layers of sizes 200 and 100 respectively. The activation function used in these layers is the rectified linear unit (ReLU). The final layer is FC, with the number of nodes given by the number of class labels and a softmax activation function to yield probabilities for each class. The optimizer used is Adam, with a learning rate $\alpha = 0.001$ and categorical cross-entropy loss function. We chose a batch size of 32.
\subsection{CNN 2D}
The 2D CNN takes the whitened spectrograms as inputs. The spectrograms centered on the samples were produced with a kaiser window ($\beta=5.6$), using multiple FFTs of $0.125$\,s with overlap. The network is composed of 3 convolutional layers with kernels of sizes (4,4), (3,3), (2,2). Max pooling (2,2) was applied after each convolutional layer. The output from the CNN architecture is fed into a FC layer of size 200 and later to a FC output layer which outputs probabilities. We used the same activation functions as in the 1D case. The optimizer used is Adam, with a learning rate $\alpha = 0.001$ and categorical cross-entropy loss function. The batch size is 32.

\subsection{LSTM}
The recurrent model takes the whitened timeseries as inputs. It is composed of 2 bidirectional LSTM layers, of dimensions 64 and 32 respectively. Spatial dropout is applied on $10\%$ of the output nodes from the first recurrent layer. The outputs is passed to 4 FC layers of size 1024, 256, 64, 32 with \textit{tanh} activation functions and finally onto a FC output layer which outputs probabilities for each class through a softmax activation function.  The optimizer used is Adam, with a learning rate $\alpha = 0.001$ and categorical cross-entropy loss function. The batch size is 256.

\begin{table}[t]
    \caption{Cumulative number of triggers over all segments analysed for each interferometer individually, along with number of coincident triggers over all three detectors.}
  \begin{center}
    \begin{tabular}{|c|c|c|c|}
      \hline
      {} &
      \multicolumn{3}{c|} {\textbf{Triggers}}\\
      \hline
      \textbf{Detector} & \textit{\, Signal \, } & \textit{Noise} & \textit{\, Total \, }\\
      \hline
      Virgo V1 & $9273 $ & $47901$ & $57174$\\
      \hline
      LIGO L1 & $10480$ & $3810$ & $14290$ \\
      \hline
      LIGO H1 & $10984$ & $4103$ & $15087$ \\
      \hline
      L1, H1, V1 & $5647$ & $675$ & $6322$ \\
      \hline
    \end{tabular}
    \label{table1}
  \end{center}
\end{table}

\section{Search and classification results} 
\label{sec:res}
\begin{figure}[!h]
\centering
    \includegraphics[width=0.5\textwidth]{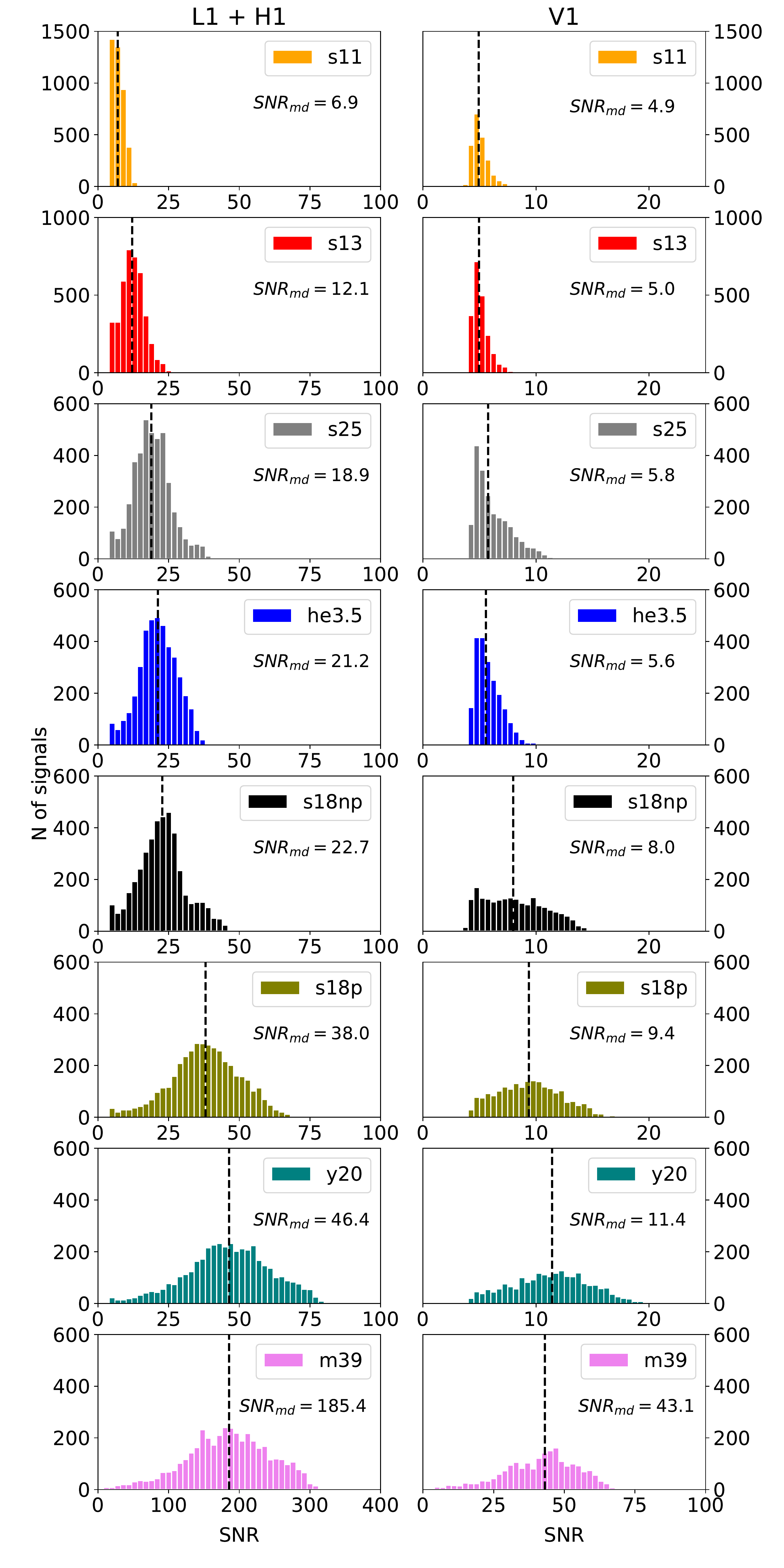}
  \caption{Matched filter S/N distribution of the signals injected at the ITFs: Virgo (\emph{right}) and LIGO Livingston and Hanford (\emph{left}). The LIGO detectors are grouped due to their similar sensitivity and signal S/N. A dashed line is drawn at the median S/N value of each waveform model.}
  \label{fig:SNR_distribution}
\end{figure}

We trained each network separately on the same training set. The full dataset was split as follows: $60\%$ training, $10\%$ validation, $30\%$ testing. In Figure \ref{fig:SNR_distribution} we show the matched filter S/N distribution of injected signals over all 44 files. The S/N of the signals varies depending on the waveform model and the simulated source position with respect to the inferferometer. As expected, the same signals produce lower S/N at Virgo due to its lower sensitivity. We observed that the matched filter S/N of lower amplitude models in Virgo amass towards an S/N equal to 4, after which the distribution drops sharply. This is consistent with the fact that we obtained the S/N as the maximum value of the matched filter response on a time series of a few seconds around each trigger. In the absence of a signal, we computed this value in the region between 3 and 4, therefore weak signals will almost never have lower S/N. Moreover, noise bursts are responsible for part of the triggers produced by WDF. In Table \ref{table1} we report the cumulative number of triggers produced over all the segments for each detector separately, along with the number of triggers coincident in the three detectors. The large excess number of noise triggers in the Virgo dataset has been obtained despite the same choice of whitening parameters when running WDF. 

This can be be explained by a non-stationary behaviour and greater variability of the detector's noise PSD  throughout the individual segments, which makes the whitening procedure less effective. This issue can be tackled in the future using adaptive whitening to track the PSD evolution in real time. For each interferometer we defined a trigger to be due to a CCSN signal when the trigger timestamp falls inside the segment covered by the injected waveform. We applied clustering when the same signal generated multiple triggers, by averaging the trigger times. Since the large number of background triggers in the V1 datum can skew the training procedure, to train this specific dataset we used only those triggers which were coincident in time with L1 and H1. The time window for coincidence was chosen to take into account the length of the waveforms and the maximum travel time between interferometers. We underline the fact that different parts of the same CCSN model can trigger WDF. Starting from the event timestamps, we chose a symmetric window around the triggers of length given by the Radice s25 model, the second shortest among the waveforms. In this way we avoided using the tails of the waveforms where the simulations are stopped due to limited computing time. We noted that the Andersen s11 waveform did not trigger the WDF pipeline a sufficient number of times in order to be effectively used during training and testing. Over the whole dataset, only 16 and 12 triggers were found coincident with s11 respectively in the LIGO H1 and LIGO L1 datasets and none coincident among all three detectors. This is the reason why we did not choose the time window for samples equal to the length of s11, while still keeping the model and its label in the training and testing procedure for completeness. 
\begin{figure}[!hb]
  \centering
    \includegraphics[width=0.50\textwidth]{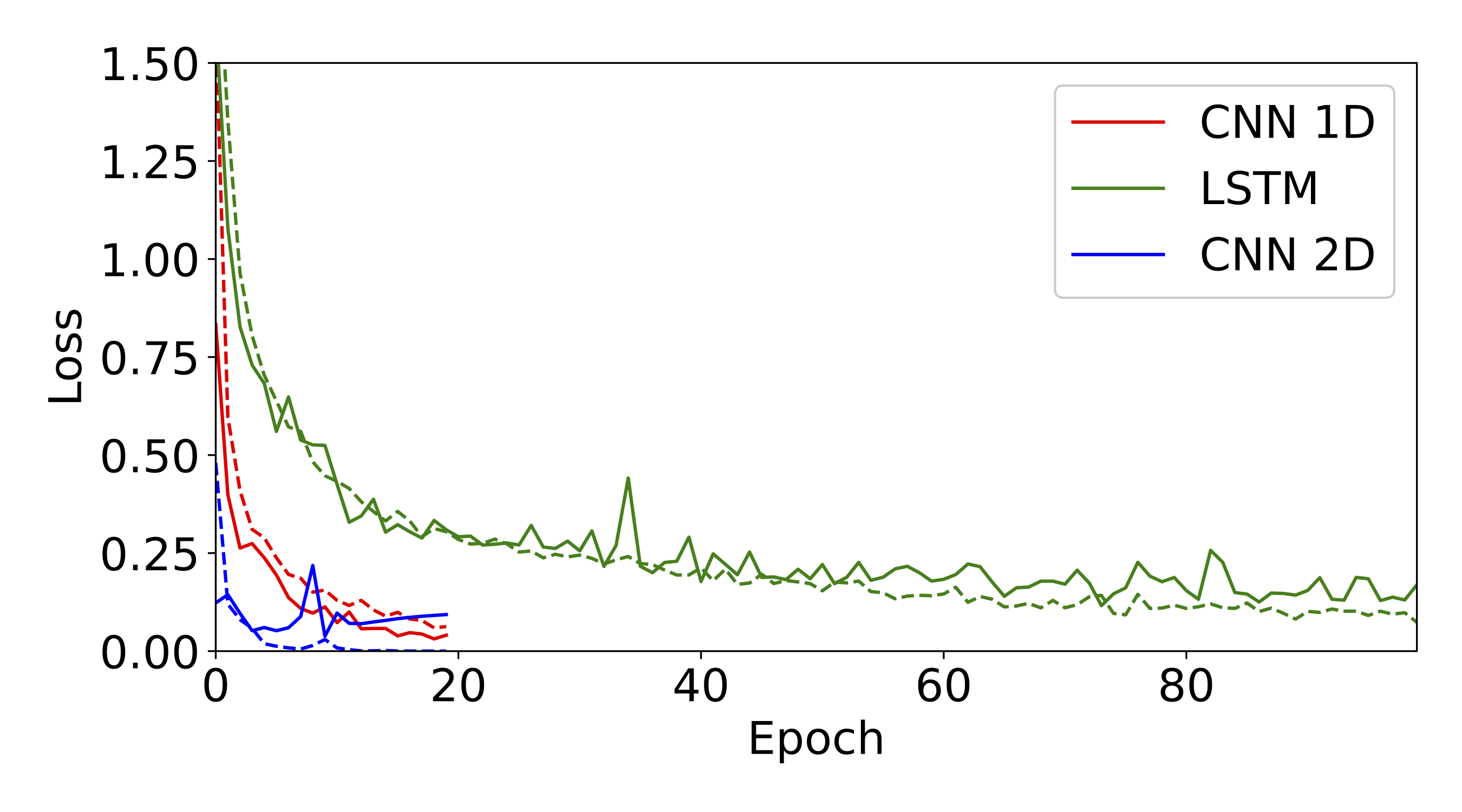}
  \caption{Evolution of the categorical cross-entropy cost function through different training epochs for the models considered. The dashed lines are training losses, while the solid
  lines are the losses computed on the validation set.}
  \label{CCE}
\end{figure} 

\begin{table*}[ht!] \tiny
    \caption{True positive rates for the noise and CCSN signal classes, along with the total sensitivity, computed for the three detectors V1, L1, H1 on real data from the O2 science run. The asterisk indicates that a model is not present in the associated dataset.}
  \begin{center}
    \begin{tabular}{|c|c|c|c|c|c|c|c|c|c|c|c|c|c}
      \hline
      {} & {} & \multicolumn{10}{|c|}{\textbf{Waveform }}\\
      \hline
      {\textbf{ITF}} & \textbf{Model} & \textit{Noise} & \textit{s11} & \textit{s18p} & \textit{s18np} & \textit{He3.5} & \textit{m39} & \textit{y20} & \textit{s13} & \textit{s25} & \textbf{Total}\\
      \hline
      \multirow{3}{*}{\textit{V1}} & LSTM & $49.2$ & * & $3.6$ & $58.4$ & $0.0$ & $89.5$ & $69.9$ & $0.0$ & $89.8$ & $73.7$ \\
      & CNN 1D & $44.6$ & * & $8.4$ & $10.9$ & $0.0$ & $84.3$ & $73.1$ & $0.0$ & $87.4$ & $68.3$ \\
      & CNN 2D & $48.6$ & * & $9.6$ & $39.4$ & $3.6$ & $92.3$ & $72.5$ & $0.0$ & $94.6$ & $75.2$ \\
      \hline

      \multirow{3}{*}{\textit{L1}} & LSTM & $90.1$ & $0.0$ & $98.2$ & $92.8$ & $85.4$ & $98.7$ & $96.0$ & $87.1$ & $94.8$ & $93.6$ \\
      & CNN 1D & $99.4$ & $0.0$ & $89.5$ & $95.3$ & $82.2$ & $99.2$ & $98.2$ & $75.5$ & $98.8$ & $95.9$ \\
      & CNN 2D & $99.8$ & $0.0$ & $99.1$ & $99.3$ & $97.4$ & $100.0$ & $99.7$ & $91.6$ & $99.8$ & $99.3$\\
      \hline

      \multirow{3}{*}{\textit{H1}} & LSTM & $96.2$ & $0.0$ & $95.5$ & $96.8$ & $89.1$ & $99.7$ & $95.9$ & $75.1$ & $97.6$ & $95.4$ \\
      & CNN 1D & $99.0$ & $0.0$ & $90.1$ & $99.3$ & $91.6$ & $98.4$ & $100.0$ & $80.6$ & $97.4$ & $96.5$ \\
      & CNN 2D & $99.7$ & $0.0$ & $99.6$ & $99.8$ & $96.8$ & $99.7$ & $99.8$ & $96.8$ & $99.2$ & $99.1$ \\
      \hline
    \end{tabular}
    \label{table2}
  \end{center}
\end{table*}
Before training, the samples were shuffled to have a uniform dataset. The time needed for training varies for each model as it is dependent on the number of weights to be optimised in their architectures. Moreover, it is related to the number of samples in each dataset and is therefore shorter for Virgo due to the use of triple coincidence. The number of epochs was separately chosen for each algorithm in order to avoid overfitting. We trained the 1D CNN on a Tesla k40 GPU for 10 epochs, averaging approximately $\sim15$\,s per epoch on L1.  The 2D CNN was trained on a Tesla k40 GPU for 10 epochs, averaging approximately $\sim20$\,s per epoch on L1. The LSTM network was trained for 100 epochs on a Tesla k80 GPU, averaging approximately $\sim60$\,s per epoch on L1, as shown in Figure \ref{CCE}. In Table \ref{table2} we show the full sensitivities of the models for all the classes at each detector, which correspond to the values in the diagonal of the confusion matrices. We recall that the sensitivity, also known as the true positive rate (TPR), to a particular class $c$, is defined as the ratio of the true positives and the sum of true positives and false negatives:

\begin{equation}
    TPR_c = \frac{TP_c}{TP_c+FN_c} \, . 
\end{equation}

\begin{figure*}[h!]
\begin{minipage}{.5\textwidth}
\centering
    \hspace*{1.0cm}
    \includegraphics[width=0.65 \textwidth]{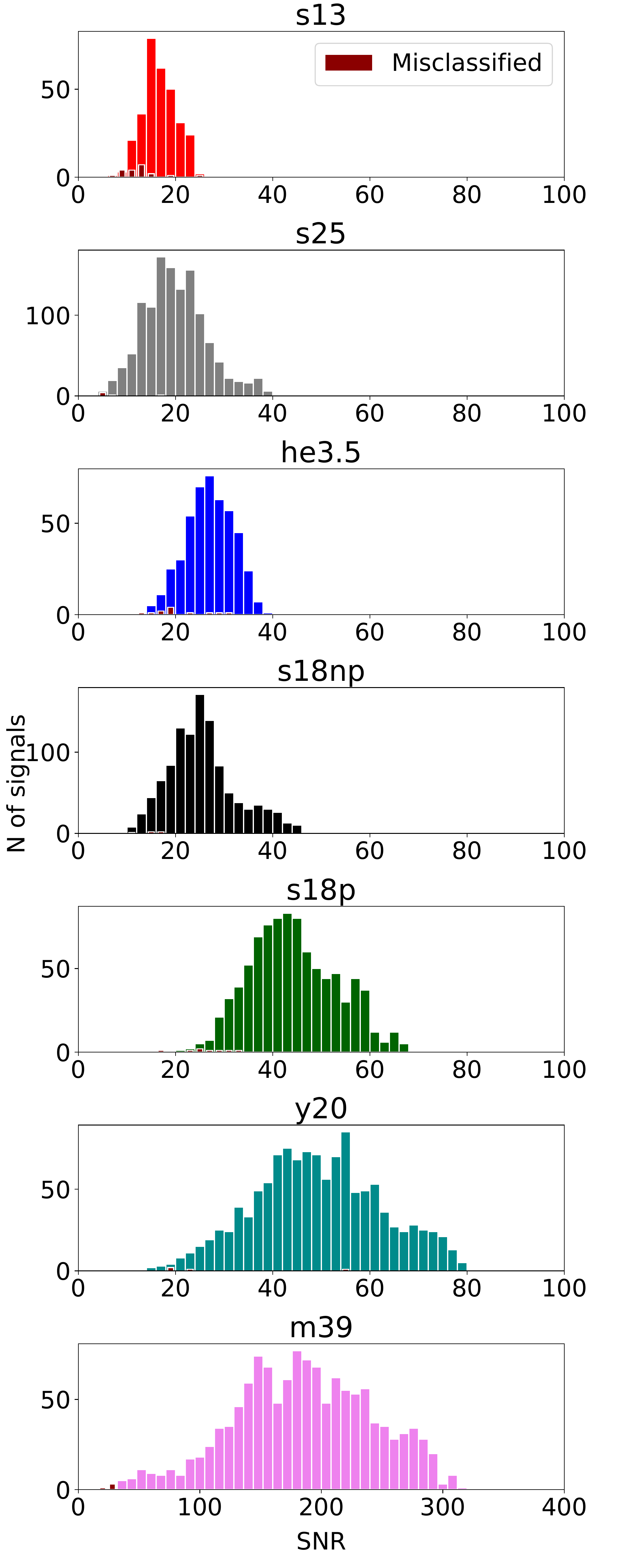}
\end{minipage}
\begin{minipage}{.5\textwidth}
\centering
    \hspace*{-1.5cm}
    \includegraphics[width=0.65 \textwidth]{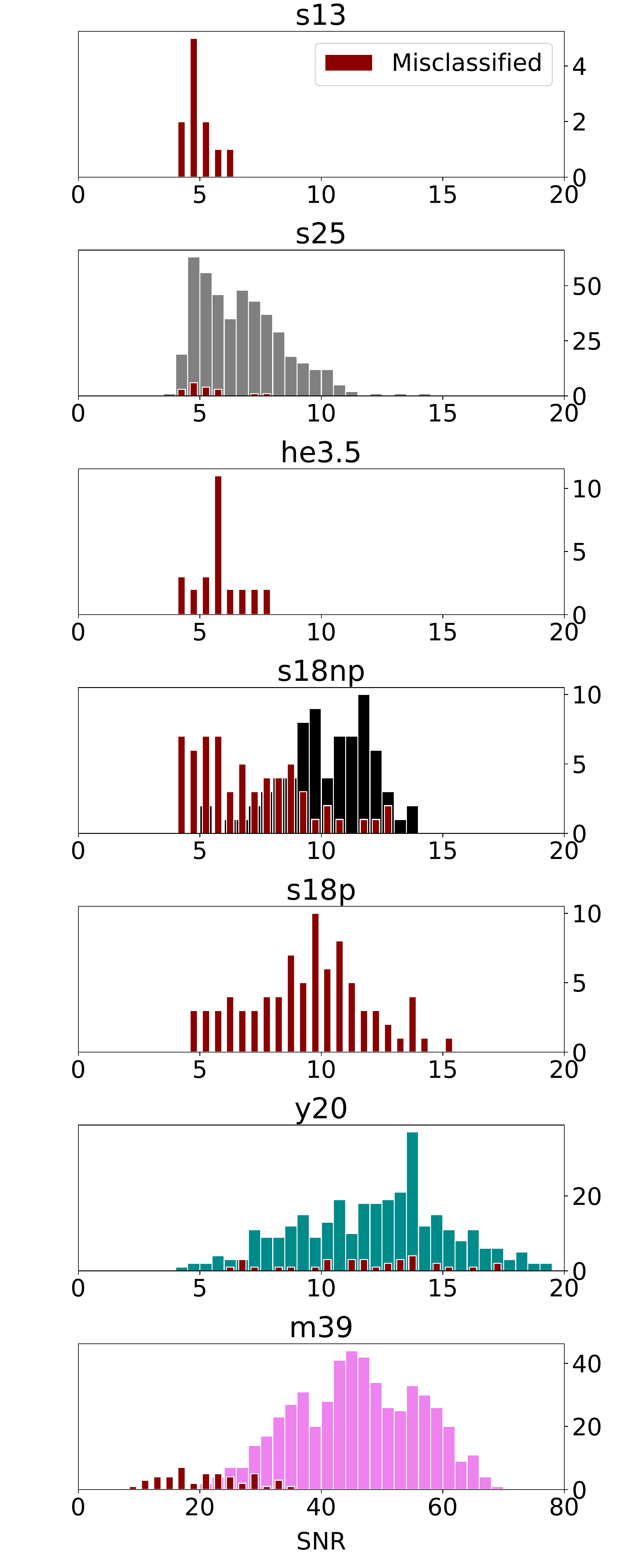}
\end{minipage}
  \caption{Histograms for matched filter S/N distribution of the LIGO L1, H1 (\emph{left}), and Virgo V1 (\emph{right}) test sets. The dark red thinner bars represent the number of signals misclassified by the merged LSTM and CNN model at each S/N interval. The bin widths are 2 for L1, H1 and 0.5 for V1, with the exception of m39 which has a larger S/N range. For the latter model the binning is equal to 8 for L1, H1 and 2 for V1.}
  \label{misc_L1V1}
\end{figure*}

We analyse each dataset in the following subsections, providing detailed results for the separate models and for the merged model. The predictions of the merged model were obtained by averaging the output probabilities computed by the 1D CNN, 2D CNN and LSTM. In Figure \ref{fig:multilabel_merged} we show the confusion matrices for the multi-label classification in all three interferometers, along with the S/N distribution of all the misclassified samples. In Figure \ref{misc_L1V1} we illustrate how the different CCSNe models were classified by the merged model in the LIGO and Virgo datasets as a function of S/N. The figure highlights the fact that CCSN signals have higher S/N in the LIGO datasets, with misclassifications occurring only for weaker signals. On the other hand, only the most energetic models injected into Virgo datum were correctly classified. An exception to this is provided by s25, which exhibits low-frequency SASI emission.

\subsection{LIGO L1}

After trigger extraction by WDF, the LIGO L1 dataset is composed in the following way: 3810 noise triggers, 12 s11, 1438 s18p, 1782 s18np, 704 he 3.5, 2052 m39, 1969 y20, 476 s13, 2047 s25. As previously described, the number of Andresen s11 samples is too low for any algorithm to train to recognise its characteristics. All ML models are more accurate in classifying the CCSN signals with higher energies, while the worst sensitivities are exhibited for he3.5 and s13, which are under represented in both training and test set compared to the rest. Specifically, the 1D CNN is less accurate with the s13 and s18p signals by $\sim10\%$ compared to the other algorithms. The LSTM has a small tendency to misclassify noise into low S/N signal classes, which may be due to the fact that it keeps memory of small noise peaks in the data. We merged the probabilities from the three models through averaging and obtained an overall sensitivity of $\sim99.3\%$, which is comparable to the best model (2D CNN). The incorrectly classified signals are at S/N values below 20.
\subsection{LIGO H1}
As expected the LIGO H1 dataset is quite similar to that of LIGO L1. Its composition is: 4103 noise triggers, 25 s11, 1534 s18p, 1803 s18np, 881 he3.5, 2059 m39, 1991 y20, 637 s13, 2054 s25. The high amplitude models achieve the best classification TPR for all the ML models used. As for L1, the s13 and he3.5 achieve the lowest accuracies for all ML algorithms. However, in this dataset the  LSTM architecture is the less accurate in classifying these signals. While still performing worse than the CNNs, the same model achieves better accuracy in the noise class compared to the L1 scenario. This may be due to the larger number of noise samples included.  Merging the output from the three models did not improve the sensitivity compared to the best model 2D CNN, but achieved similar values ($TPR\sim99.1\%$). As for L1, lower S/N, below 20, contribute to most of the misclassified samples.

\subsection{Virgo V1}
The triple-coincident triggers between V1, L1, and H1, which are used to identify samples in the Virgo V1 dataset, are distributed as follows: 675 noise, 0 s11, 329 s18p, 491 s18np, 115 he3.5, 1940 m39, 1139 y20, 76 s13, 1557 s25. The lower amplitude CCSN models are either not present at all (s11) or appear in a small number of samples, predominantly at lower S/N. None of the ML algorithms considered were able to effectively classify s13, he3.5, and s18p signals.\begin{figure*}[h!]
\begin{minipage}{.6\textwidth}
\centering
    \vspace*{-0.5cm}
    \includegraphics[width=0.7 \textwidth]{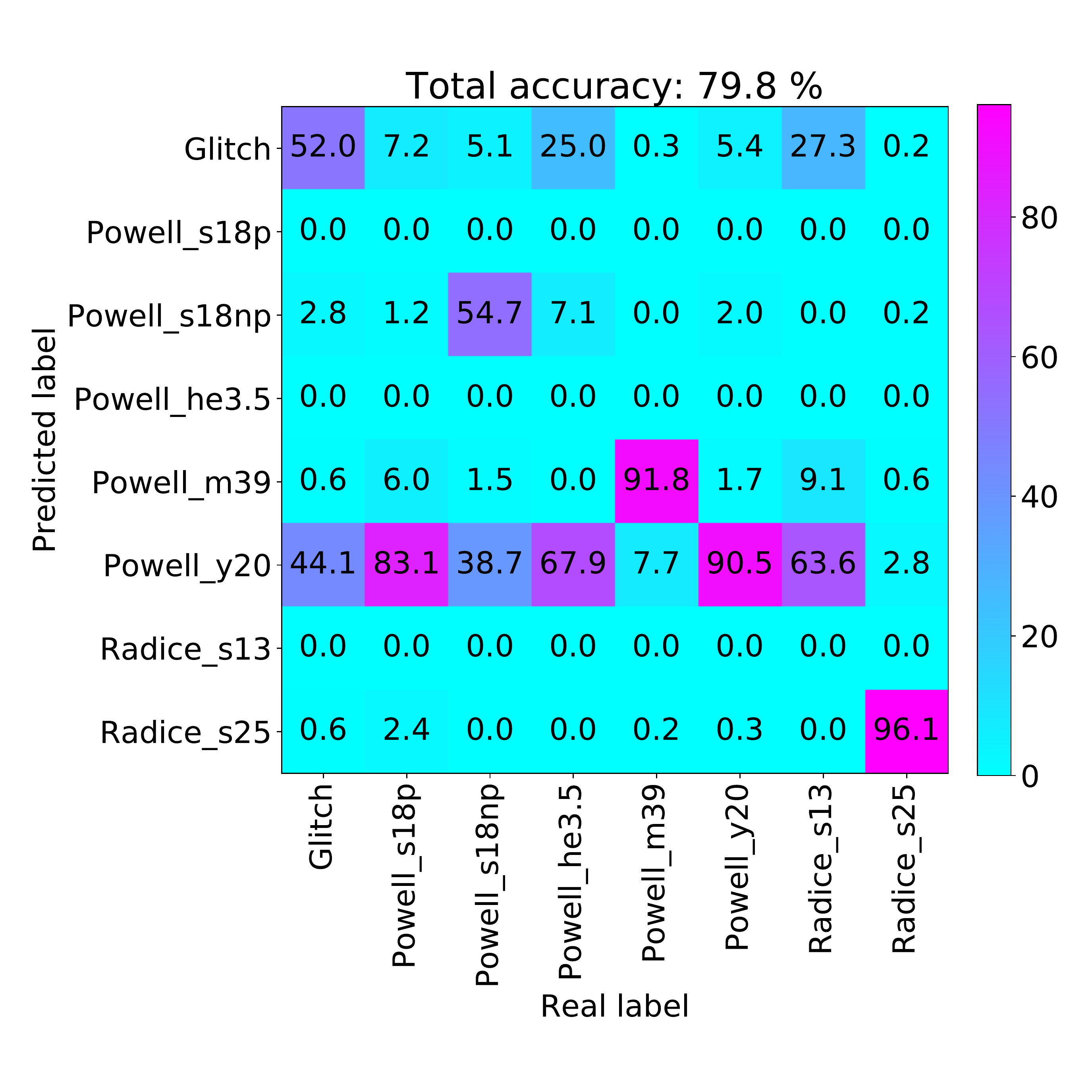}
\end{minipage}
\begin{minipage}{.8\textwidth}
\centering
    \vspace*{-1.5cm}
    \hspace*{-8cm}
    \includegraphics[width=0.5 \textwidth]{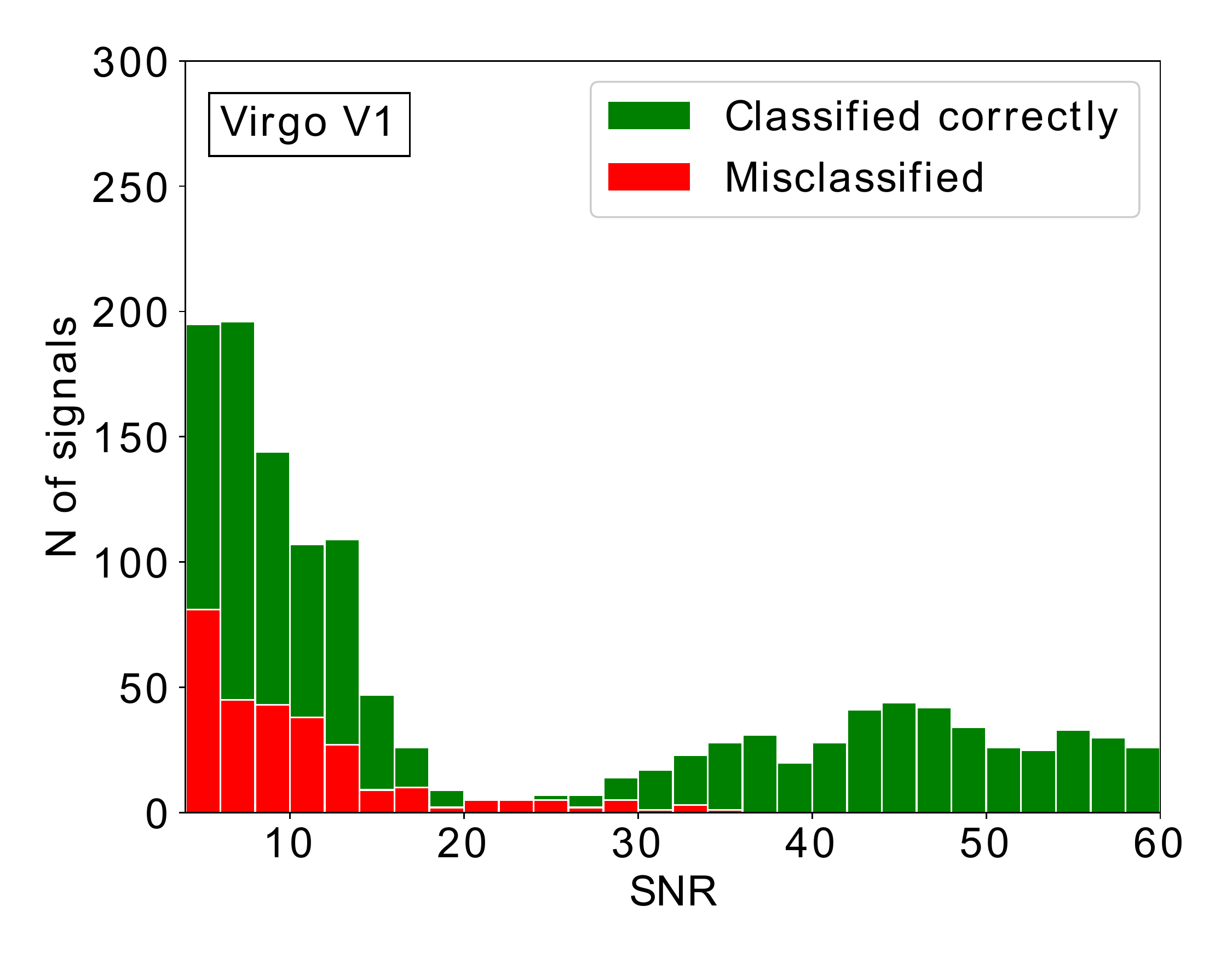}
\end{minipage}
\begin{minipage}{.6\textwidth}
\centering
\vspace*{-0.5cm}
    \includegraphics[width=0.7 \textwidth]{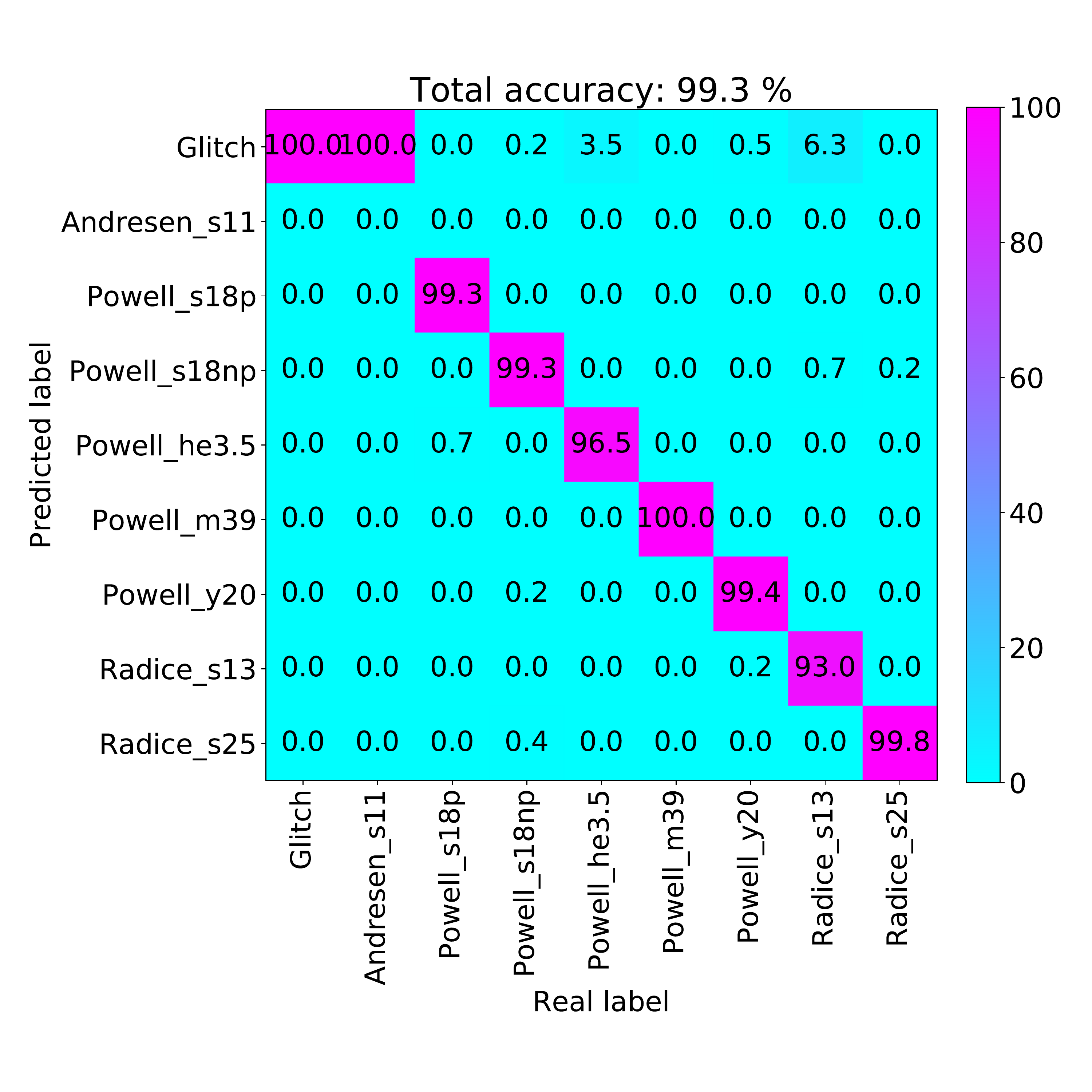}
\end{minipage}
\begin{minipage}{.8\textwidth}
\centering
    \vspace*{-1.5cm}
    \hspace*{-8cm}
    \includegraphics[width=0.5 \textwidth]{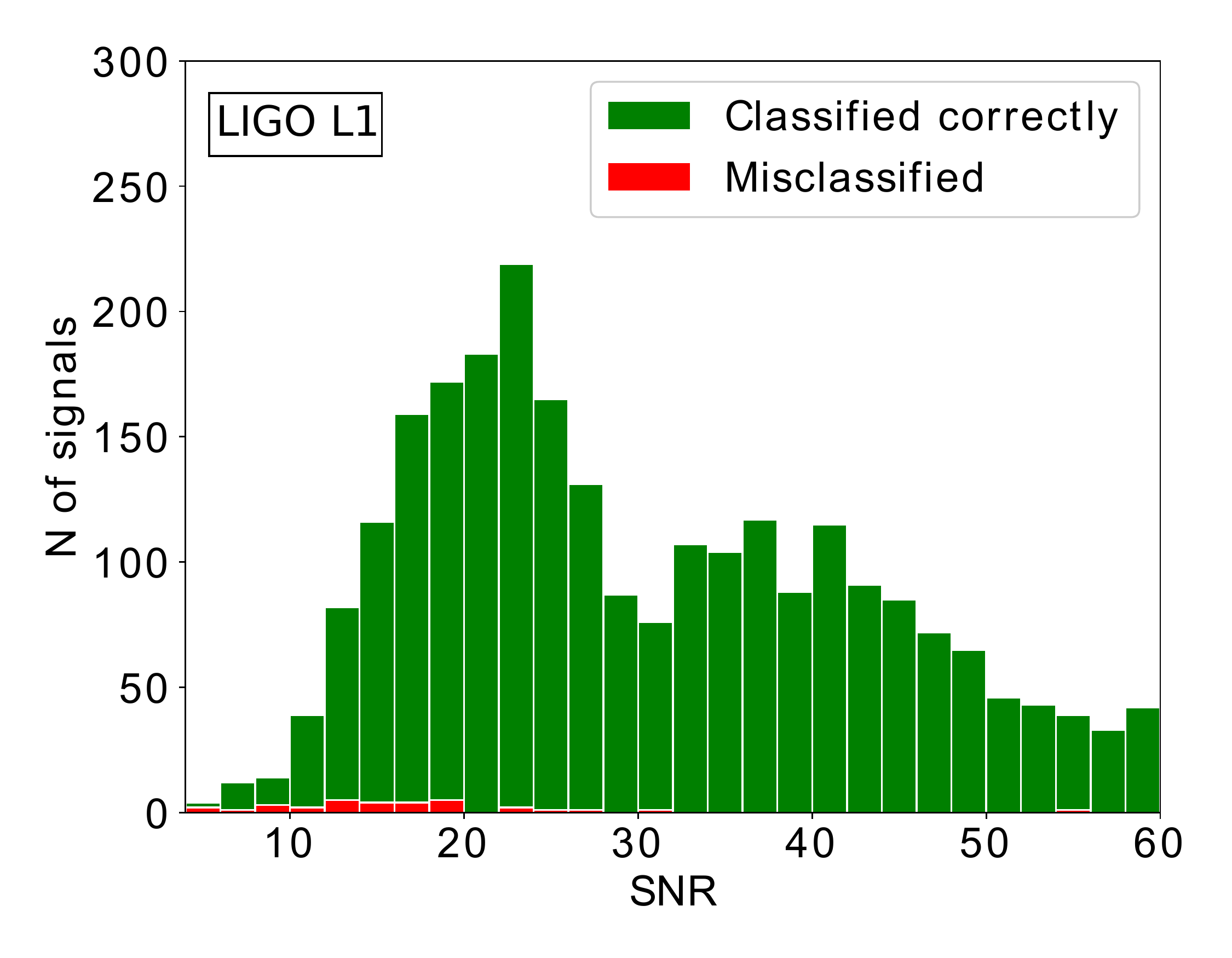}
\end{minipage}
\begin{minipage}{.6\textwidth}
\centering
    \vspace*{-0.5cm}
    \includegraphics[width=0.7 \textwidth]{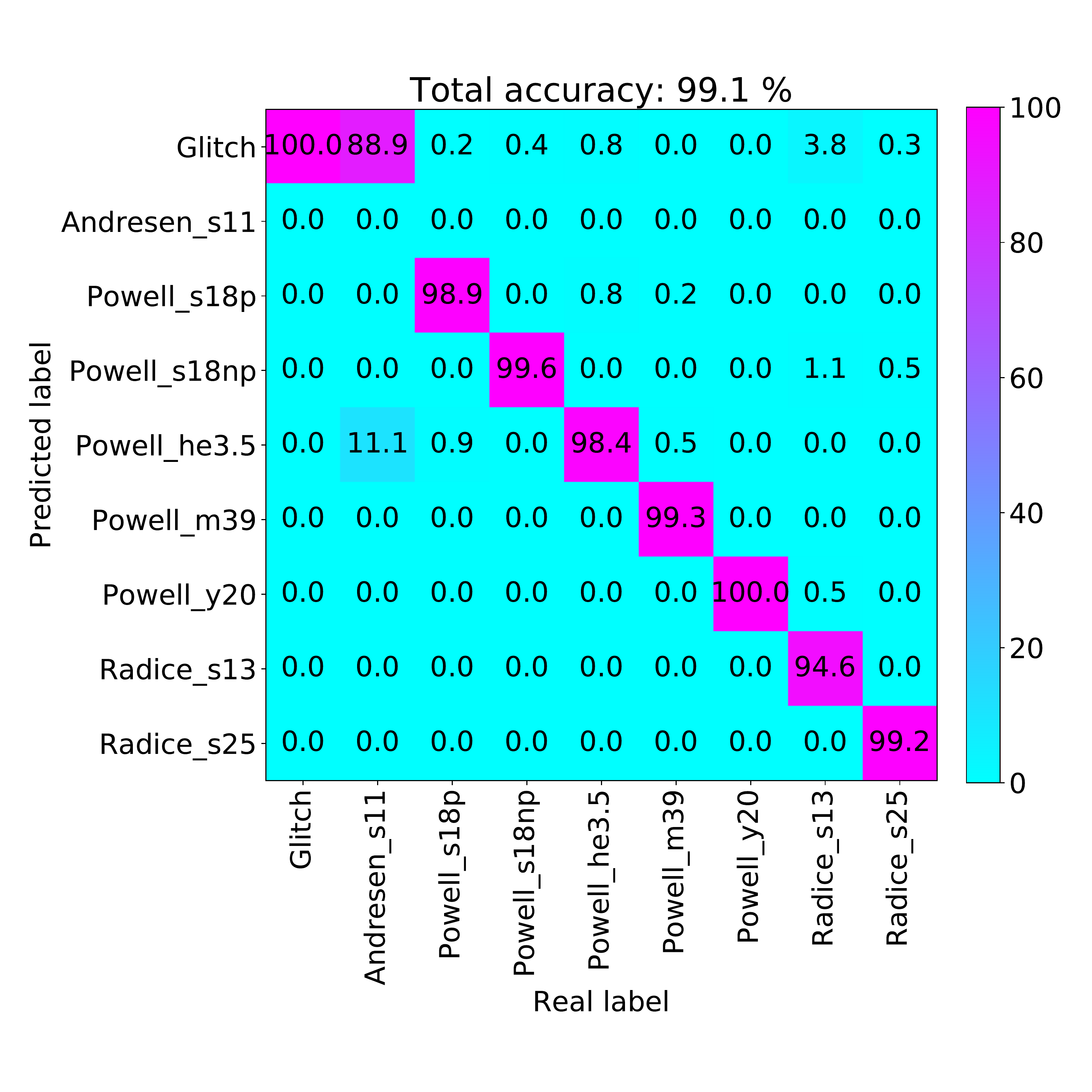}
\end{minipage}
\begin{minipage}{.8\textwidth}
\centering
    \vspace*{-1.5cm}
    \hspace*{-8cm}
    \includegraphics[width=0.5 \textwidth]{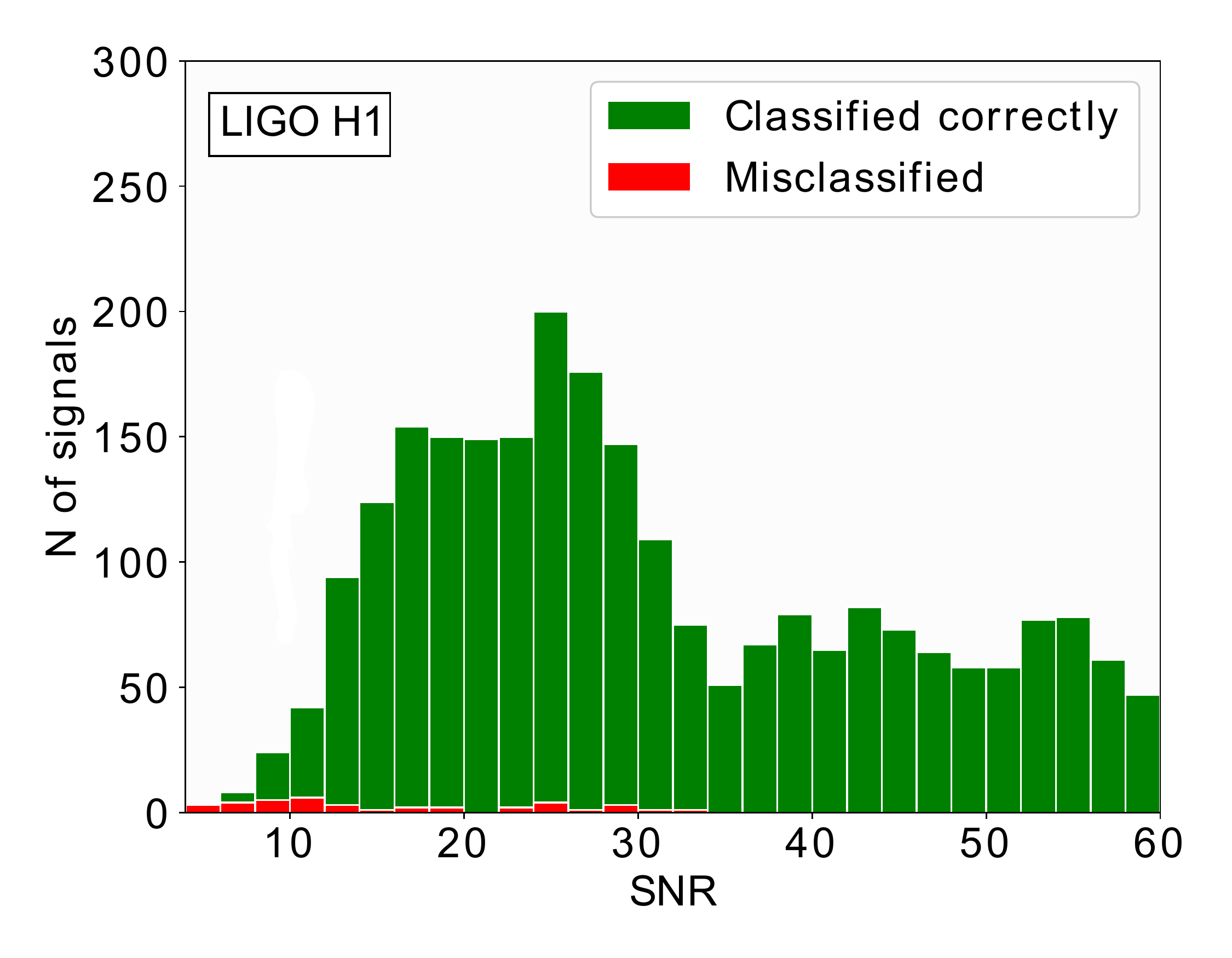}
\end{minipage}
  \caption{Multi-label classification confusion matrices for the V1 (\emph{top}), L1 (\emph{middle}), and H1 (\emph{bottom}) datasets using the merged output from LSTM, 1D CNN, and 2D CNN, along with the S/N distribution of samples classified correctly and incorrectly. We show only the S/N range below S/N=60, where misclassifications occur.}
  \label{fig:multilabel_merged}
\end{figure*}
 The LSTM network fares better than the CNNs for the mid-amplitude s18np model which presents a low frequency SASI emission. A possible explanation is that the LSTM is able to recognize the low frequency emission at lower S/N. The LSTM performs slightly worse than 1D CNN and 2D CNN on the y20 test samples. All models achieve good sensitivities ($TPR>84\%$) on the m39 and s25 models. The most common misidentification is with noise triggers at low S/N and with the y20 model. For V1 data the merged output improves the overall sensitivity with respect to the 2D CNN, with a $TPR\sim79.8\%$ compared to $\sim75.3\%$. As evident from Figure \ref{fig:multilabel_merged}, lower S/N<20 contribute to most of the misclassified samples, although weaker m39 signals misidentifications occur up to $S/N\sim30$. The S/N distribution of the signals exhibits a gap in the $S/N=18-30$ region: this is due to the different intrinsic GW amplitudes of the CCSN models used in the analysis and as a consequence leads to most signals being misclassified in the corresponding S/N interval. 

\subsection{Three interferometer analysis}
\begin{figure}[!ht]
  \centering
    \includegraphics[width=0.5\textwidth]{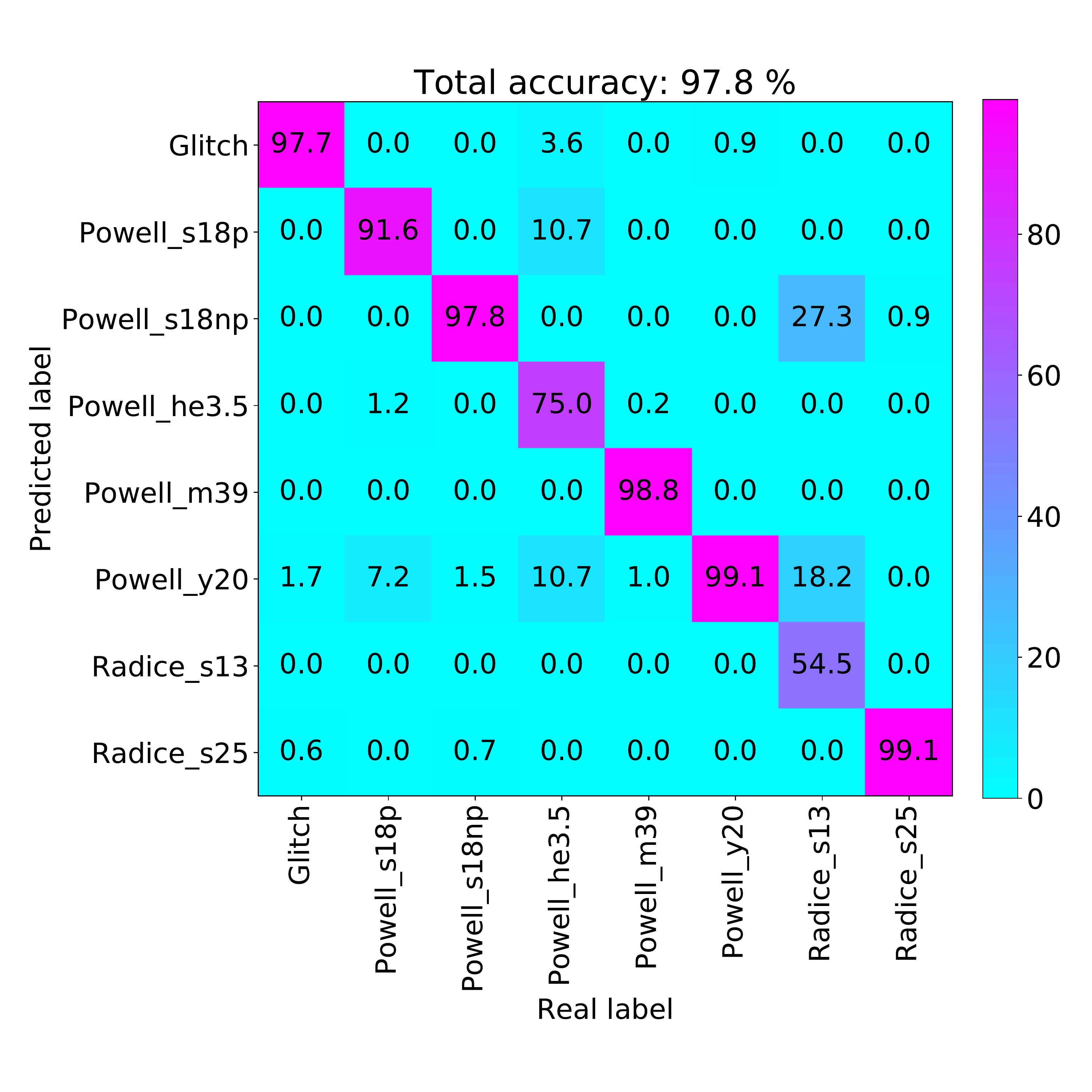}
  \caption{Test set classification confusion matrix obtained by merging the outputs of LSTM, 1D CNN, and 2D CNN models trained on samples from a three interferometer network with LIGO (L1, H1) and Virgo (V1) data.}
  \label{3ITF}
\end{figure}

While we managed to obtain good single detector classification accuracies for the LIGO interferometers for signals at higher S/N, a more robust and realistic approach is to exploit data from all available detectors to ideally increase the accuracy for weaker signals. This can be achieved training a network on three detector data by stacking the individual time series or images, increasing the number of dimensions of the inputs by one. The CCSN signals are defined using time and equatorial coordinates in the geocentric frame. They are then projected into each interferometer datum taking into account the antenna pattern and the time shifts between the detectors. We trained and tested the models on the three interferometer samples obtained using the coincident triggers previously computed for the Virgo V1 dataset. We refer to that dataset for a breakdown of the signals used for each category. As a result the sample size is much smaller than the single LIGO detectors cases and this negatively affects the performance of the classifier. Despite such caveat, the sensitivities of the merged model to the individual CCSN models pictured in Figure \ref{3ITF}, are still very high and comparable to those obtained for the individual LIGO detectors. In detail, we see a familiar pattern where the under-represented classes and lower S/N models (he3.5, s13) have lower accuracies. On the other hand, the classification sensitivities are significantly higher than those of the Virgo dataset for all classes. As in the single detector case, the merged model is more robust in correctly classifying noise samples with a $TPR\sim97.7 \%$ in the noise class; this statement holds despite the fact that the noise class amounts to a much smaller percentage of the overall dataset, as evident in Table \ref{table1}. The overall $TPR$ for the merged model is $\sim97.8\%$. The TPR computed over all classes for the individual models confirms the single detector results and is, respectively, $TPR\sim89.5\%$ for the 1D CNN, $TPR\sim86.7\%$ for the LSTM, and $TPR\sim98.4\%$ for the 2D CNN. 
Since the reduced number of triple-coincident samples hampers the training procedure, a possible solution to increase the classification accuracy is to rely on the use of generative adversarial networks (GANs) \citep{goodfellow2014generative} to augment the dataset. GANs have already been successfully applied in the generation of realistic, synthetic datasets for the variety of images of people \citep{gulrajani2017improved} or even fingertips \citep{yu2019attributing}. The same approach could in principle improve balance between various CCSN models in the studied datasets for the spectrogram representation. However, that approach would require a separate, dedicated research project.

\section{Discussion} 
\label{sec:disc}
Among the sources which emit GWs that have yet to be observed, CCSNe are specifically interesting due to their multiple emission channels that convey different informations on the collapse mechanisms in play. The current matched filter techniques cannot be applied for a GW search for CCSNe, due to the large degree of stochasticity in the emitted signal. ML algorithms are largely agnostic about data and do not make assumptions on the signal morphology. However, they are well suited at detecting patterns in timeseries and time-frequency data. We have previously proven the efficacy of 1D and 2D CNNs in classifying signals based on their morphology in both domains, assuming a gaussian noise background. In this follow-up study we proved that both CNN architectures and also recurrent LSTM networks can obtain good accuracies also on classification tasks involving real detector noise. Compared to \cite{Iess2020}, this analysis presents important differences. Firstly, we added three new neutrino-driven explosion models to increase the complexity of the multi-label classification problem. Secondly, we grouped real noise transients in a single class during the training procedure. Moreover, we fixed the distance of all signals to $1$\,kpc and carried our single detector analysis on the same sources at each of the three interferometers, to compare S/N distributions and classification accuracies. As a result, the signal S/N will only depend on the sensitivities of the detectors at injection time and on the direction in the sky relative to the detector plane, through the antenna pattern functions. All three networks achieved accuracies above $90\%$ on the LIGO datasets and the predictions were consistent for the two detectors with similar sensitivity and orientation. As in the case of simulated noise background, the 2D CNN showed the best performance among the three and was very stable in its prediction for separate training runs. 1D CNN performed slightly better than LSTM for the LIGO datasets, but worse in the Virgo dataset, largely due to its inability to classify the s18np model which presents a SASI signature. The performance of the merged model is comparable to the 2D CNN, with a very high accuracy in distinguishing noise instances when fed with a large number of samples. For the Virgo dataset accuracies were expected to be worse due to the smaller sample size and the lower S/N of the CCSNe signals at the detector. Nonetheless higher amplitude signals generally achieved moderately good accuracies. On the other hand, results at S/N$<15$ for the same dataset are strongly affected by the sample size. To fully assess this low S/N region, an ad hoc dataset with a larger number of samples for each waveform can be built to probe the performance at such S/N. In this study we also extended our previous analysis \citep{Iess2020} to a multi-detector framework as in \cite{2018PhRvD..98l2002A} and \cite{PhysRevD.102.043022} by stacking the samples for a given signal from each interferometer, requiring time-coincidence in the WDF triggers. This allowed us to perform the analysis on a realistic scenario in which the LIGO datasets contain higher S/N CCSN samples compared to the Virgo interferometer. We achieved slightly lower accuracies than the single LIGO interferometer case due to the smaller dataset arising from requiring triggers to be coincident. However, we predict that increasing the number of samples in the dataset will eventually fill the gap between the two cases. From a computational point of view we are limited by dataset size and GPU memory. While we achieved similar accuracies with LSTM and 1D CNN architectures, hyper-parameter tuning was harder for the recurrent model, which also needed significantly longer training times. Moreover, LSTMs decrease their performances when faced with the task of learning longer sequences, a fact that limits their capabilities when analyzing datasets which are composed of thousands of timesteps. The better performance of LSTMs compared to 1D CNNs in the lower S/N region for some models which exhibit a low frequency emission component requires further analysis with a larger dataset. Downsampling may also be implemented to enhance the computational performance of the recurrent model and reduce the length of the sequence to be learnt. Overall, the above analysis suggests that convolutional and recurrent NN architectures can play a role in GW detection pipelines for neutrino-driven CCSN burst signals, with the added possibility of distinguishing different types of signal morphologies for sources located in a nearby region of the Milky Way. The results obtained are consistent with applications to neutrino-driven CCSN models on simulated \citep{Iess2020} and real detector noise \citep{Portillo2021}, at the fixed distance range considered in this study. Past and current estimates have set the rate of galactic CCSN between 1 and 2 events per century (\cite{1991ARA&A..29..363V}, \cite{1993A&A...273..383C}, and \cite{Rozwadowska2021}). Taking into account planned improvements in the sensitivity of the Virgo, LIGO and KAGRA interferometers, a galactic CCSN detection with deep learning methods seems possible if an event occurs with a GW signature that shares some characteristics (g-mode emission, prompt convection, SASI) with current simulations.

\section*{Acknowledgments}
\label{sec:ack}
This article/publication is based upon work from COST Action CA17137, supported by COST (European Cooperation in Science and Technology).
FM is supported by the Polish National Science Centre grants 2016/22/E/ST9/00037, 2017/26/M/ST9/00978, and 2020/37/N/ST9/02151 as well as Polish National Agency for Academic Exchange grant PPN/IWA/2019/1/00157. OL and CN are supported by the STFC UCL Centre for Doctoral Training in Data Intensive Science grant ST/P006736/1 and by an STFC consolidated grant number ST/R000476/1. Part of the computations were carried out on the PL-GRID Prometheus cluster infrastructure. This research has made use of data, software, and/or web tools obtained from the Gravitational Wave Open Science Center (https://www.gw-openscience.org/ ), a service of LIGO Laboratory, the LIGO Scientific Collaboration, and the Virgo Collaboration. LIGO Laboratory and Advanced LIGO are funded by the United States National Science Foundation (NSF) as well as the Science and Technology Facilities Council (STFC) of the United Kingdom, the Max-Planck-Society (MPS), and the State of Niedersachsen/Germany for support of the construction of Advanced LIGO and construction and operation of the GEO600 detector. Additional support for Advanced LIGO was provided by the Australian Research Council. Virgo is funded, through the European Gravitational Observatory (EGO), by the French Centre National de Recherche Scientifique (CNRS), the Italian Istituto Nazionale di Fisica Nucleare (INFN), and the Dutch Nikhef, with contributions by institutions from Belgium, Germany, Greece, Hungary, Ireland, Japan, Monaco, Poland, Portugal, and Spain. The data that support the findings of this study are available from the corresponding author upon reasonable request.

\bibliographystyle{aa}
\bibliography{bibfile}

\end{document}